\documentclass{article}

\usepackage{graphicx}
\usepackage{url}

\newcommand{\eq}[1]{(\ref{#1})}
\newcommand\ff[1]{#1}

\title{Characteristics of latitude distribution of sunspots and their links to solar activity in pre-Greenwich data}
\author{V.G.~Ivanov and E.V.~Miletsky\\[1ex]
Central Astronomical Observatory at Pulkovo,\\
Saint-Petersburg, Russia\\[1ex]
E-mail: vg.ivanov@gao.spb.ru
}
\begin{document}

\maketitle

\begin{abstract}
We study and compare characteristics of sunspot group latitude
distribution in two catalogs: the extended Greenwich (1874--2014)
and Schwabe ones (1825--1867) \cite{arlt13}. We demonstrate that both
datasets reveal similar links between latitude and amplitude
characteristics of the 11-year cycle: the latitude dispersion
correlates with the current activity and the mean latitude of
sunspots in the cycle's maximum is proportional to its amplitude, It
agrees with conclusions that we made in papers
\cite{imn,im14} for
the Greenwich catalog.

We show that the latitude properties of sunspot distribution are
much more stable against loss of observational data than traditional
amplitude indices of activity. Therefore, the found links can be
used for estimates of quality of observations and independent
normalizing of activity levels in a gappy pre-Greenwich data. We
demonstrate it using the Schwabe catalog as an example.

In addition, we show that the first part of the Schwabe data
probably contains errors in determination of sunspot latitudes that
lead to overestimation of the sunspot latitude dispersions.
\end{abstract}

\section{Introduction}

To characterize the level of solar activity one traditionally uses
amplitude indices, which are calculated on the base of number and
size of sunspots (the Wolf number, Group Sunspot Number, sums of
sunspot group areas etc). However, until the beginning of the epoch
of regular observations of sunspots those indices are often derived
from non-uniform data, which contain errors due to loss or incorrect
treating o of observations. For example, it is known (see, e.g.,
\cite{vit}), that the widely used Z\"urich series of
the Wolf number before the middle of the 19th century was
constructed by R. Wolf on the base of a fragmentary data.

In addition to amplitude indices there are data on spatial
distributions of sunspot groups. First of all, such data are
presented in the Greenwich catalog of sunspot groups. Recently other
catalogs with sunspot coordinates for earlier epochs became
available, e.g., the catalogs based on observations of Staudacher
\cite{arlt09} and Schwabe \cite{arlt13}.

On the one hand, both information on number of sunspots and on their
latitude distribution is subjected to distortions caused by loss of
observational data, but it is much weaker for the latter. On the
other hand, there are stable links between the latitude distribution
of sunspots in the 11-year solar cycle and its amplitude
\cite{imn,im11,im14}. Therefore, latitude characteristics of
sunspots can be used for control and correction of normalization of
traditional series of amplitude indices. In this paper we
demonstrate it analyzing the extended Greenwich catalogue (GC),
that include the original Greenwich data and their extension by
NOAA/USAF (\url{http://solarscience.msfc.nasa.gov/greenwch.shtml}), and
the Schwabe catalog (SC) \cite{arlt13}
(\url{http://www.aip.de/members/rarlt/sunspots/schwabe}).

\section{Data and method}

It is convenient to use as an amplitude characteristic of solar
activity the index G that is equal to yearly averaged daily numbers
of observed sunspot groups. This index can be readily obtained from
sunspot groups catalogs, it is tightly related to the Group Sunspot
Number index (GSN) proposed by Hoyt and Schatten \cite{hoyt} and
differ from the latter basically in its normalization (G $\approx$
GSN/12). As a measure of the sunspot latitude extension we will use
the yearly means of absolute values of sunspot group latitudes $\phi$
and their dispersions averaged over the two hemispheres
$\sigma_{\phi}^2$.

\begin{figure}
\begin{center}
\ff{\includegraphics[width=0.99\textwidth]{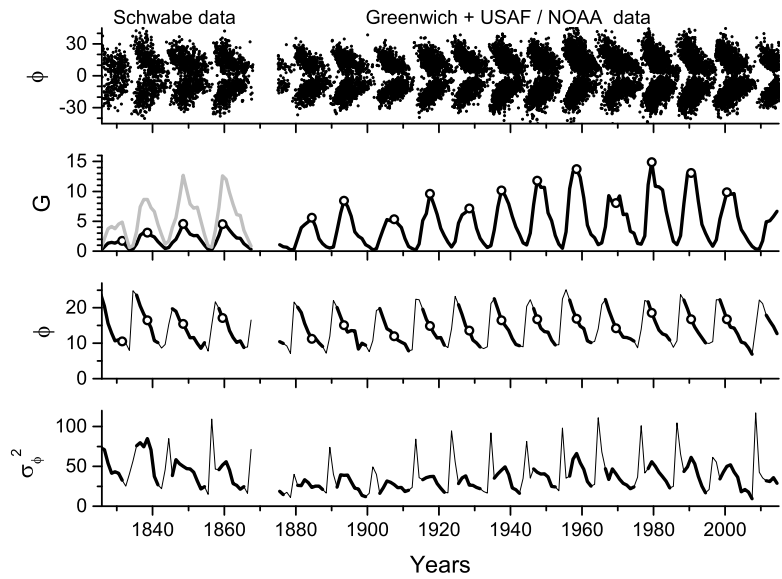}}
\caption{Top panel: the latitude distribution of sunspot groups
(``the Maunder butterflies'') for  GC (1874--2014) and SC
(1825--1867).  Lower panels: indices G, $\phi$ and $\sigma_{\phi}^2$
for both catalogs. See the text for details.} \label{fig1}
\end{center}
\end{figure}

In Fig.~\ref{fig1} the latitude distribution of groups (``the
Maunder butterflies'') and indices  G, $\phi$ and $\sigma_{\phi}^2$
for GC (1874--2014) and SC (1825--1867) are plotted. In SC each
drawing of the Sun is attributed by the so called ``subjective
quality flag'' $Q$, and in the following we will use, unless
otherwise stated, only data with $Q=1$ corresponding to the highest
quality. The thin lines in plots of $\phi$ and $\sigma_{\phi}^2$
correspond to years of cyclic minimums and two adjacent years, which will
not be taken into account in analysis of latitude properties, since
in these years the wings of neighboring Maunder butterflies overlap,
so the mean latitudes are ambiguous and the hemisphere dispersions
are strongly overestimated.

\begin{figure}
\begin{center}
\includegraphics[width=0.99\textwidth]{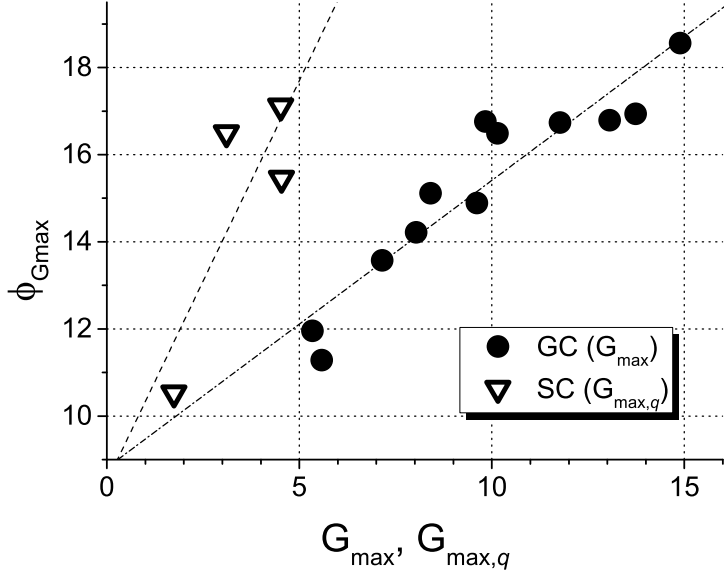}
\caption{The relationship  between amplitudes of 11-year cycles ${\rm
G}_{\rm max}$ and the mean latitudes in maximums $\phi_{{\rm G}_{\rm
max}}$ for GC (circles) and the similar relationship  for ``raw'' indices
$G_{{\rm max},q}$ in SC (triangles).} \label{fig2}
\end{center}
\end{figure}

The empty circles in Fig.~\ref{fig1} mark maximums of index G and
the mean latitudes $\phi_{G_{\rm max}}$ corresponding to these
moments. (for the bimodal 20th cycle the moment between two almost
equal peaks is selected as a maximum). The values $G_{\rm max}$ and
$\phi_{{\rm G}_{\rm max}}$ are well correlated  (the correlation
coefficient $r=0.93$, see Fig.~\ref{fig2}) and related by the
regression equation
\begin{equation}\label{eq1}
\phi_{{\rm G}_{\rm max}}=0.66^\circ\cdot {\rm G}_{\rm max}+8.81^\circ \,,
\end{equation}

For the Wolf numbers a similar relationship was found by Waldmeier as early
as in the 1930s \cite{wald39,wald55}. It is tightly connected to the
following two regularities:

(i) Evolution of the mean latitude of sunspots in the 11-year cycle
(``the Sp\"orer law''), as it was demonstrated by Ivanov and Miletsy
in \cite{im14}, can be described by the universal
dependence $\phi(t) = A\cdot \exp \left[-b \cdot (t - T_{\rm min})
\right]$, where $T_{\rm min}$ is the moment of the cycle minimum,
the coefficient $a$ correlates with the amplitude of the cycle and
$b \approx -0.13\,{\rm years}^{-1}$ does not depend upon this
amplitude;

(ii) According to the Waldmeier rule \cite{wald35} maximums in
higher cycles tend to take place earlier than in lower ones.

One can easily deduce from rules (i) and (ii) that the mean latitude
in the maximum must be higher for more powerful cycles, in agreement
with expression \eq{eq1}.

\begin{figure}
\begin{center}
\includegraphics[width=0.99\textwidth]{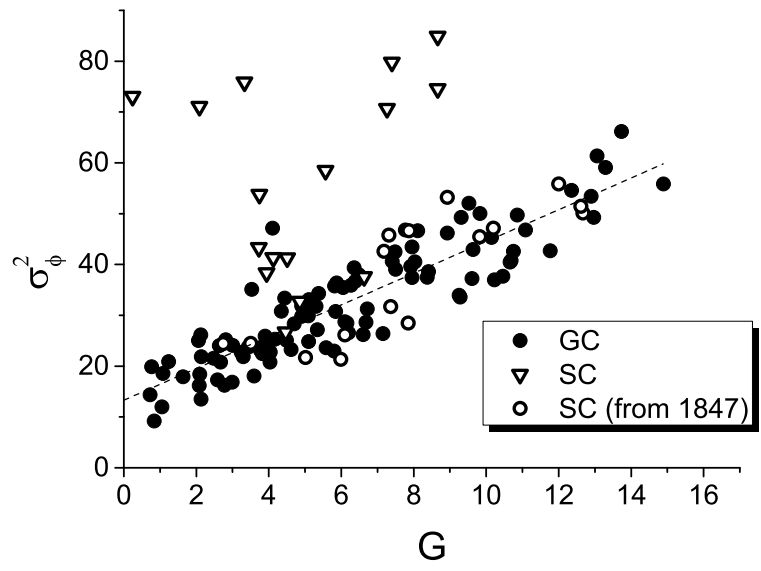}
\caption{The relationship
between the index G  and the latitude dispersion $\sigma_\phi^2$ for
GC (filled circles), entire SC (triangles) and SC from 1847 (empty
circles).}
\label{fig3}
\end{center}
\end{figure}

Besides, in papers \cite{mi09,imn,im11} we found a relationship between the
latitude extension of the sunspot distribution and the level of solar
activity. For GC such relationship between G and the dispersion
$\sigma_\phi^2$ (see Fig.~\ref{fig3}) can be described by the regression
\begin{equation}\label{eq2}
\sigma_\phi^2=3.12 \cdot {\rm G} + 13.3\;{\rm deg}^2
\end{equation}
with correlation $r = 0.90$.

\begin{figure}\begin{center}
\includegraphics[width=0.99\textwidth]{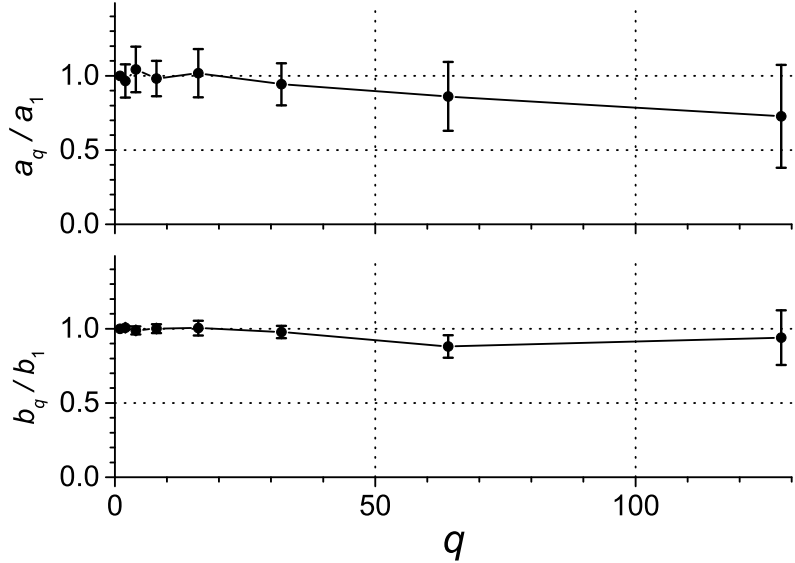}
\caption{Behavior
of ratios $a_q/a_1$ and $b_q/b_1$, which characterize regressions
\eq{eq1s} and \eq{eq2s}, for the sparsed GC as a function of the loss
factor $q$. The error bars for each point are estimated
by 100 randomly sparsed series
} \label{fig4}
\end{center}
\end{figure}

It is important to note that the both mentioned relations are not
destroyed in cases when a part of observations is lost. To show it,
we artificially sparsed GC, randomly selecting from the data
one qth part of observations. Regressions \eq{eq1} and \eq{eq2} in
this case turn to
\begin{equation}\label{eq1s}
\phi_{{\rm G}_{{\rm max},q}} = a_q \cdot q \cdot {\rm G}_{{\rm max},q} + c_q
\end{equation}
and
\begin{equation}\label{eq2s}
\sigma_{\phi,q}^2 = b_q \cdot q \cdot {\rm G}_q + d_q\,,
\end{equation}
where variables with indices $q$ correspond to the sparse GC and the
additional ``loss factor'' $q$  in ${\rm G}_q$ and ${\rm G}_{{\rm
max},q}$ compensates the loss of $(q-1)/q \cdot 100\%$ data in the
catalog. Behavior of the ratios $a_q/a_1$ and $b_q/b_1$, which
characterizes change of the relationships between amplitude and latitude
extension of sunspot activity with growth of $q$ as compared with
the same relationships for the full GC (i.e. for $q=1$), is presented in
Fig.~\ref{fig4}. One can see that even for $q = 100$ (i.e. when 99\%
of observations are lost), relative variations of the regression
coefficients are limited by the range 20--25\%.

Such stability of the found relationships against loss of data allows their using
for control or/and restoration of normalization of amplitude indices
of solar activity in catalogs of sunspots.

\section{Restoration of normalization in the Schwabe catalog}

Let us demonstrate how relation \eq{eq1} can be used for restoration
of amplitude indices by the example of the Schwabe catalog. To do it
we plot for SC the dependence ${\rm G}_{{\rm max},q} - \phi_{{\rm
G}_{\rm max}}$ (Fig.~\ref{fig2}) and build the corresponding
regression
\begin{equation}\label{eq3}
\phi_{{\rm G}_{\rm max}}
 = 1.84^\circ \cdot {\rm G}_{{\rm max},q} +8.49^\circ \qquad (r=0.82)\,,
\end{equation}
where ${\rm G}_{{\rm max},q} $ are maximums of cycles for the
``raw'' indices ${\rm G}_q$, calculated by SC and the coefficient $q
>1$ corresponds to some ({\em a priori} unknown) loss of data. Comparing \eq{eq1}, \eq{eq1s}
and \eq{eq3}, one can find $q = 1.84^\circ / 0.66^\circ \approx 2.8$
and obtain ``renormed'' indices ${\rm G} = q \cdot {\rm G}_q$ (the
gray curve in the second panel of Fig.~\ref{fig1}) with the
distortion due to the data loss corrected.

\begin{figure}
\begin{center}
\includegraphics[width=0.99\textwidth]{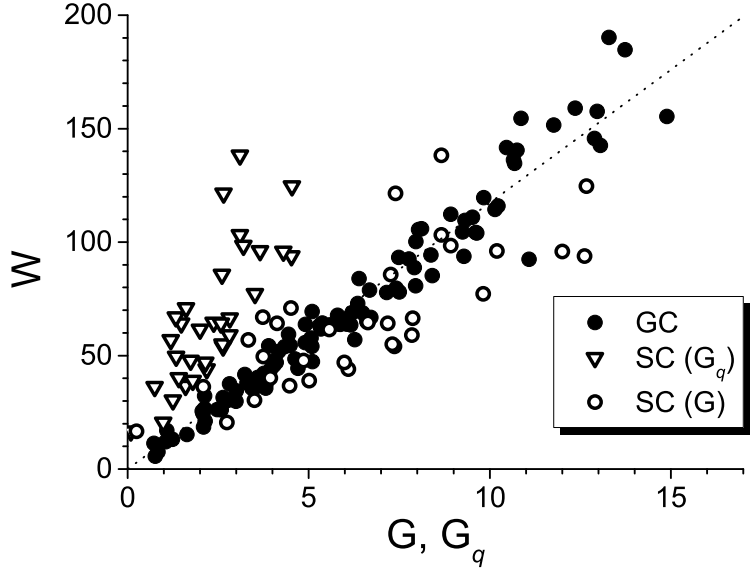}
\caption{The relation between indices W and G for GC (filled
circles), ``raw'' ${\rm G}_q$ (triangles) and renormed G (empty
circles) indices for SC.} \label{fig5}
\end{center}
\end{figure}

In this case we can independently control agreement of the obtained
renormalization using the Wolf numbers W that are known for this
epoch. In Fig.~\ref{fig5} the relations between indices W and G  for
GC, ``raw'' (${\rm G}_q$) and renormed (G) indices for GC are shown.
One can see that the normalization, which we obtained without using
of the known Wolf numbers, agrees with the latter rather well.

\section{Latitude dispersions in the Schwabe catalog}

Let us study the second relationship (2) between G and $\sigma_\phi^2$
(Fig.~\ref{fig3}), using the renormalized G found for SC. One can
see that the properties of this dependence vary and it agrees with the
relation found for GC only since 1847. If we assume that the
normalization of index G for SC is valid, we conclude that until the
middle of the 1840s the dispersions of the latitude distribution are
anomalously large. One can make the same conclusion from
Fig.~\ref{fig1}, where in the corresponding cycles the unusually
great number of sunspot are sited close to the equator.

\begin{figure}
\begin{center}
\includegraphics[width=0.99\textwidth]{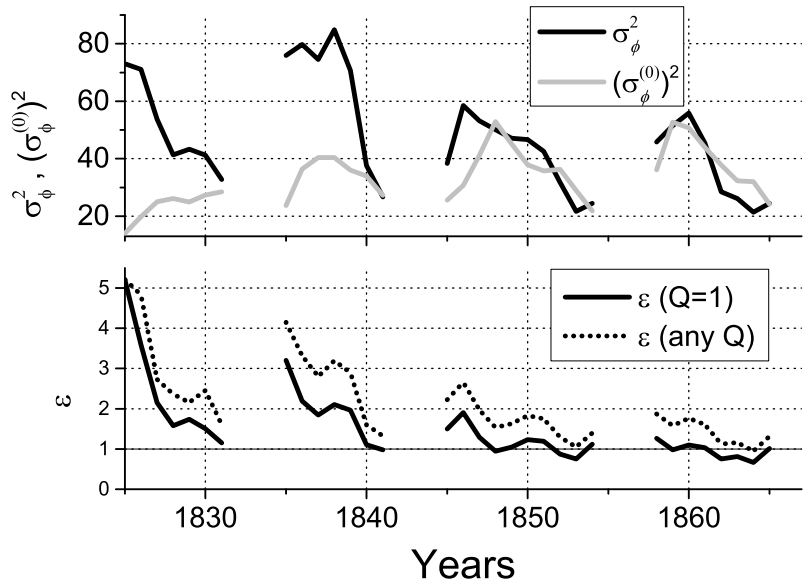}
\caption{Top panel: the observed latitude dispersions
$\sigma_\phi^2$ for SC (black curves) and the corresponding values
$(\sigma_\phi^{\rm (0)})^2$ obtained by the relation (2) (gray
curves). Bottom panel: overestimatings of the dispersion in SC
$\epsilon = \left( \sigma_\phi /\sigma_\phi^{\rm (0)} \right)^2$ for
$Q=1$ (solid curves) and any Q (dotted curves). } \label{fig6}
\end{center}
\end{figure}

In the top panel of Fig.~\ref{fig6} the latitude dispersions
$\sigma_\phi^2$, which are calculated  directly from the data of SC,
and the values $(\sigma_\phi^{\rm (0)})^2$ obtained from G with use
of the relation (2), are plotted. We assume that the relation (2)
remained valid for sunspots in the pre-Greenwich epoch, so the
visible difference in the dependence between activity and latitude
dispersion of GC and SC are caused by some errors in the latter.
This error can be described by the value $\epsilon = \left(
\sigma_\phi /\sigma_\phi^{\rm (0)} \right)^2$ (see the bottom panel
of Fig.~\ref{fig6}).  This relation is large for the cycles~7 and 8
and decreases almost to unity after the middle of 1840s. Let us note
that if one uses all data of SC (the dotted curves of the bottom
panel) rather than observations with the quality flag $Q=1$ only
(the solid curves), the magnitude of $\epsilon$s become notably
larger. Therefore, we assume that GC contains errors in
determination of sunspot latitudes that are larger for drawings of
low quality and decrease to the end of the period of observations.
It seems probable that the cause of these errors is wrong
determination of the position of the solar equator in drawings
of Schwabe. Apparently, in the case of systematic inclinations of
the equator line on the drawings of the Sun relative to its true
angle the calculation of the latitude distribution must lead to
overestimation of $\sigma_\phi^2$. At the same time, it will not
cause a systematic shift of the mean latitudes $\phi$ because of
equal probabilities of positive and negative errors in the
inclination. Just such a picture one can see in the data of SC.

It is interesting that a similar anomalous number of sunspots on the
equator were found in the observations of Staudacher (see
Fig.~2 in the paper \cite{arlt09}). Arlt assumes that such
phenomenon can be caused by a quadrupole magnetic field dominating
on the Sun in the third quarter of the 18th century. However, the
same picture can be a result of errors in determination of the
equator position on the drawings of the observer.

\section{Conclusions}

Therefore, relationships between characteristics of the latitude
distribution of sunspots and the level of solar activity allow one
to control normalization of activity indices and correct their
distortions caused by a loss of a part of data. In this paper we
discuss two such relationships. The first one \eq{eq1} relates the mean
latitude of sunspots in the maximum of activity $\phi_{{\rm G}_{\rm
max}} $ with the amplitude of the 11-year cycle ${\rm G}_{\rm max}$.
The second relationship (2) associates the dispersion of the mean latitude
$\sigma_\phi^2$ and the current level of activity G.

Apparently, evaluation of the mean latitudes requires less precision
in determination of sunspot coordinates than calculation of the
latitude dispersions. We saw it on the example of SC,  the first
part of which, probably, contains large errors. On the other hand,
using of the relationship \eq{eq1} requires that a catalog does not contain
dramatic changes in data quality and the loss factor $q$ does not
vary strongly during a 11-year cycle. At the same time, usage of the
relationship (2) does not limited by the data quality so strongly, since
it operates by yearly indices. In cases when using of both
relationships is possible they, as it was shown above by the example of the
second part of SC, lead to consistent results.

It is interesting that both relationships hold even in the case of loss of
99\% of data (see Fig.~\ref{fig4}), i.e. in situations when a
direct calculation of amplitude indices (like the Wolf number)
becomes very difficult. This fact makes possible using the described
latitude-amplitude relations for analysis and correcting of solar
activity indices obtained on the base of pre-Greenwich sunspots
catalogs.

\section{Acknowledgements}

The paper was supported by the RFBR grant No. 13-02-00277 and
programs of the Presidium of the Russian Academy of Sciences Nos.~21
and~22.

\end{document}